\documentclass[prd,superscriptaddress,showpacs,nofootinbib,amsmath,amssymb,aps,10pt]{revtex4}

\usepackage{bm}
\usepackage{amsfonts}
\usepackage{latexsym}
\usepackage[latin1]{inputenc}
\usepackage{graphicx}
\usepackage{amsmath}
\usepackage{palatino}
\usepackage{mathpazo}
\usepackage{epstopdf}
\usepackage{booktabs}
\usepackage{dcolumn}
\usepackage{array}

\usepackage{wrapfig}
\usepackage{wasysym}

\def\jnl@style{\it}
\def\aaref@jnl#1{{\jnl@style#1}}

\def\aaref@jnl#1{{\jnl@style#1}}

\def\aj{\aaref@jnl{AJ}}                   
\def\apj{\aaref@jnl{ApJ}}                 
\def\apjl{\aaref@jnl{ApJ}}                
\def\apjs{\aaref@jnl{ApJS}}               
\def\apss{\aaref@jnl{Ap\&SS}}             
\def\aap{\aaref@jnl{A\&A}}                
\def\aapr{\aaref@jnl{A\&A~Rev.}}          
\def\aaps{\aaref@jnl{A\&AS}}              
\def\mnras{\aaref@jnl{Mon.~Not.~Roy.~Astron.~Soc.}}             
\def\prd{\aaref@jnl{Phys.~Rev.~D}}        
\def\prc{\aaref@jnl{Phys.~Rev.~C}}  
\def\prl{\aaref@jnl{Phys.~Rev.~Lett.}}    
\def\qjras{\aaref@jnl{QJRAS}}             
\def\skytel{\aaref@jnl{S\&T}}             
\def\ssr{\aaref@jnl{Space~Sci.~Rev.}}     
\def\zap{\aaref@jnl{ZAp}}                 
\def\nat{\aaref@jnl{Nature}}              
\def\aplett{\aaref@jnl{Astrophys.~Lett.}} 
\def\apspr{\aaref@jnl{Astrophys.~Space~Phys.~Res.}} 
\def\physrep{\aaref@jnl{Phys.~Rep.}}      
\def\physscr{\aaref@jnl{Phys.~Scr}}       
\def\commat{\aaref@jnl{Comm.~Math.~Phys.}}              
\def\science{\aaref@jnl{Science}}               
\def\cqg{\aaref@jnl{Classical Quant.~Grav.}}            
\def\jpcs{\aaref@jnl{JPCS}}                                     
\def\ijmpd{\aaref@jnl{Int.~J.~Mod.~Phys.~D}}                    
\def\grg{\aaref@jnl{Gen.~Relat.~Gravit.}}               
\def\rpp{\aaref@jnl{Rep.~Prog.~Phys.}}          
\def\npa{\aaref@jnl{Nucl.~Phys.~A}}        
\def\lrr{\aaref@jnl{Living Rev.~Rel.}}                   
\def\jcap{\aaref@jnl{J.~Cosmology Astropart.~Phys.}}    
\def\rmp{\aaref@jnl{Rev.~Mod.~Phys.}}   

\allowdisplaybreaks[1]

\begin{document}

\title{Moment of inertia -- compactness universal relations in scalar--tensor theories and $\mathcal{R}^2$ gravity}

\author{Kalin V. Staykov}
\email{kstaykov@phys.uni-sofia.bg}
\affiliation{Department of Theoretical Physics, Faculty of Physics, Sofia University, Sofia 1164, Bulgaria}

\author{Daniela D. Doneva}
\email{daniela.doneva@uni-tuebingen.de}
\affiliation{Theoretical Astrophysics, Eberhard Karls University of T\"ubingen, T\"ubingen 72076, Germany}
\affiliation{INRNE - Bulgarian Academy of Sciences, 1784  Sofia, Bulgaria}

\author{Stoytcho S. Yazadjiev}
\email{yazad@phys.uni-sofia.bg}
\affiliation{Department of Theoretical Physics, Faculty of Physics, Sofia University, Sofia 1164, Bulgaria}
\affiliation{Theoretical Astrophysics, Eberhard Karls University of T\"ubingen, T\"ubingen 72076, Germany}


\begin{abstract}

We are investigating universal relations between different normalisations of the moment of inertia and the compactness of neutron and strange stars. Slowly rotating as well as rapidly rotating models are studied in General Relativity, $\mathcal{R}^2$ gravity and scalar--tensor theories of gravity.
Moment of inertia -- compactness relations are examined for different normalisations of the moment of inertia. It is shown that for all studied cases the deviations from EOS universality are small for the examined equations of state. It turns out that in some of the cases  the examined relations are also theory independent to a good extent. Universality in relations between the maximum mass and the moment of inertia for some unstable models is also investigated.

\end{abstract}
\maketitle
\date{}
\section{Introduction}

General Relativity (GR) is very thoroughly tested in the weak field regime and the results undoubtedly confirm its predictions. However, this is not the case for the strong field regime and there are concerns regarding the viability of the theory  in this case. The recently confirmed accelerated expansion of the universe is a major issue in GR. There are two ways of solving that problem -- one is to introduce a new type of matter with exotic properties that interacts only gravitationally with the visible matter. This is the so-called dark energy and it is supposed to constitute around 73 \% of the total amount of mass and energy in the universe. Another way to deal with this is to suggest that we do not have complete knowledge for the gravitational theory so we should study alternative and modified theories.

We are in the beginning of a new era in GR testing. The new generation of gravitational wave detectors and  radio astronomy observatories, both ground based and space based, will give us a unique possibility for testing the strong field regime with neutron stars and black hole candidates. In the context of this it is crucial to test extensively  different viable alternative theories of gravity. In this paper we continue our  studies of two such alternative theories.

A class of such theories are the so-called $f(\mathcal{R})$ theories: the Lagrangian of the General Relativistic Hilbert-Einstein action, that is equal to  the Ricci scalar $\mathcal{R}$, is exchanged with a more general one. The new Lagrangian is a function of $\mathcal{R}$, hence $f(\mathcal{R})$ theories \cite{Sotiriou2010,DeFelice2010a}.

Another class are the so-called scalar--tensor theories (STT). In this case in addition to the tensorial character of GR there is a scalar field \cite{Fierz56,Jordan59,brans611,Damour1992,Will1993,Will2006}. In this class of theories a spontaneous scalarization might occur for specific initial conditions. In this case the STT solutions are energetically more favourable than the GR ones.

Due to the uncertainty in the equation of state (EOS), finding universal relations (independent from the EOS) between the star's parameters is quite appealing. One of the first hints for such relations can be found in the work of Anderson and Kokkotas \cite{Andersson98a}. In this work the fundamental and the pressure mode frequencies of an oscillating neutron star as functions of the average density $\sqrt{M/R^3}$ and the compactness $M/R$ are investigated. Some empirical relations between the parameters of the star are derived. This study has been expanded on the last two decades.  Other realistic EOS  have been included \cite{Benhar04} and later other relations have been constructed that can be quite EOS independent. In a recent work Lau et al. \cite{Lau2010} exchanged the compactness with the so-called "effective compactness", $\eta \equiv \sqrt{M^3/I}$. The purpose of this is a  better EOS independence. In \cite{Chirenti2015, Staykov2015} the above mentioned relations are  investigated and their EOS independence is commented.

In the recent years a new type of universal relations between the normalised moment of inertia, $\bar{I}$, the Love numbers, and the normalised quadrupole moment, $\bar{Q}$, was discovered by Yagi and Yunes \cite{Yagi2013, Yagi2013a}. These relations were thoroughly investigated in the last few years in many papers \cite{Maselli2013, Doneva2014,Pappas2014, Chakrabarti2014,Haskell2014, Urbanec2013, Majumder2015, Yagi2015}. Some not so common relations like the ones in \cite{AlGendy2014a} can be found in the literature too.

Different universal relations were also studied in alternative theories of gravity  \cite{Yagi2013, Yagi2013a,Sham2014,Kleihaus2014,Pani2014, Doneva2015,Staykov2015}. It is interesting to point out that in these studies the results are not only to some degree EOS independent but in some alternative theories the results do not differ significantly from the GR ones. In this case one can speak of theory independence. From one point of view such relations can be quite useful.They can supply us with unique values for the parameter of the neutron stars which depend neither on the EOS nor on the theory. On the other hand the universal relations are one of the most promising ways for distinguishing the alternative theories of gravity from the GR case. Because of the EOS uncertainty, the dispersion of the results due to the different EOS in one theory can exceed the difference between the theories. Therefore, if we have an EOS independent relation, with difference between the alternative theories big enough to be measured, and if we can measure independently  the star's parameters we can use it to test the theory.

Empirical relations between star's parameters, as a sequence of the universal relations,  were proposed in many papers \cite{Andersson98a, Lattimer2005, Lattimer2001}. In \cite{Andersson98a} a scheme for estimating the neutron star's parameters by such empirical relation is suggested.  In \cite{Lattimer2005} the authors suggested using such a relation between the moment of inertia $I$, the mass $M$, and the radius $R$ of a neutron star. It is possible to measure $I$ and $M$ with  good accuracy, but this is not the case for the radius. Having a universal relation will allow one to obtain the radius of the neutron star with high accuracy, just knowing the other two parameters. In a recent paper Breu and Rezzolla \cite{Breu2016} thoroughly investigate such relations and comment on their applications.  The main point of their study is concentrated on the universality for models with "critical mass",  but they investigated EOS independent relations between the normalised moment of inertia and the compactness too.  A full discussion on the applications of the universal relations and a procedure for determining the radius and the radius error is incorporated in their paper. In our study we extend part of their work in the above mentioned alternative theories of gravity. We concentrate mainly on the normalised moment of inertia--compactness relations and on the models with "critical mass".

An interested reader can find a more extensive review of the different universal relations known in the literature, and how they can be used to test GR and alternative theories of gravity, in a subchapter devoted to the subject in the recent review paper by Berti et al. \cite{Berti2015}.

The structure of this paper is as follows. In Section II we comment  both on alternative theories as well as on the chosen EOS. In section III the numerical results are presented. In its first subsection  the results concerning universal relations between the moment of inertia, the mass and the radius of a neutron star are studied. In the second subsection we comment on the universality for some unstable models. The paper ends with conclusions.

\section{Analytical  basis}

In the present study we are employing two alternative theories of gravity, namely scalar-tensor theory of gravity and $f(R)$ gravity. The $f(R)$ theories are mathematically equivalent to the scalar-tensor theories  and we will exploit that equivalence in this work. This allow us to use the same set of field equations for both cases. The first one is an STT with no potential, and the second one is equivalent to STT with a non vanishing potential. More thorough mathematical explanations, the field equations explicit form and a discussion on the proper boundary conditions one can find in \cite{Yazadjiev2014, Staykov2014}. We should point out that although our work is conducted in Einstein frame and the results are transformed to Jordan frame, we double-checked our results by using and a Jordan frame code.

In the STT case  we adopt  the conformal factor

\begin{equation}
A(\varphi) = e^{\beta \varphi^2/2},
\end{equation}
where $\varphi$ is the scalar field, and $\beta$ is the free parameter of the theory.  Spontaneous scalarization is observed for $\beta \leq -4.35$ \cite{Harada1998}  in the static case and  binary pulsar observations constrain the parameter to $\beta \geq -4.5$ \cite{Will2006,Antoniadis2013}. In the present work we adopt $\beta = -4.5$.

The explicit form of the Lagrangian for $f(\mathcal{R})$ gravity is $f(\mathcal{R}) = \mathcal{R} + a\mathcal{R}^2$, the so-called $\mathcal{R}$--squared gravity. The free parameter $a$ is restricted by the observations to $ a \lesssim 5 \times 10^{11} \rm{m}^2 $ \cite{Naef2010} or in dimensionless units -- $a \sim 10^{5}$. The connection between the dimensional and the dimensionless parameter is  $a \rightarrow a/r_{g}^2$, where $r_g = 1.47664 \rm km$ is one half of the gravitational radius of the Sun. In this study we adopt $a = 10^4$, which gives  deviation from GR close to the maximal \cite{Yazadjiev2014,Staykov2014}.

In this paper we are studying slowly rotating as well as rapidly rotating neutron and strange stars. For the former ones we are using the slow rotation approximation and are solving the corresponding system of ordinary differential equations numerically. In the latter case we are using a modification \cite{Doneva2013, Yazadjiev2015} of the $RNS$ code \cite{Stergioulas95} to solve the system of partial differential equations. Although the $RNS$ code can be used for slowly rotating models, we
are using both codes to double-check our results.

\begin{figure}[]
	\centering
	\includegraphics[width=0.5\textwidth]{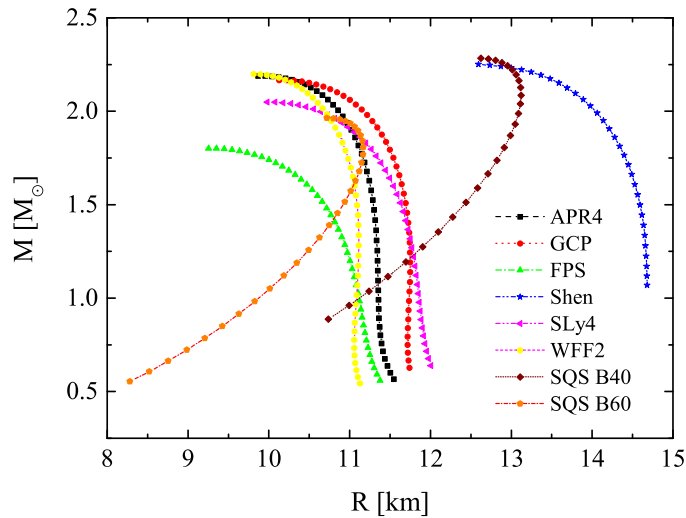}
	\caption{Mass of radius relations for all  EOS in GR. }
	\label{Fig:MR}
\end{figure}

In Fig. \ref{Fig:MR} we are plotting the mass of radius relations for all EOS we are using in this study. We work with six hadronic EOS, APR4, GCP, FPS, Shen, Sly4, WFF2, and two quark ones. The hadronic ones are tabulated and cover different stiffnesses. A special comment on EOS FPS and EOS Shen should be made. The first one we choose because in GR it gives maximal mass below the observational limit of $2 M_{\odot}$, but in $f(\mathcal{R})$ gravity the maximal mass increases above that boundary \cite{Yazadjiev2014}. We want to investigate the effect of the soft EOS on the universality and whether it will change in  different theories.  The latter EOS is often used in merging simulations and supernova collapse simulations and it also gives large radii so we find it interesting to investigate whether adding it will break the universality.
The quark EOS has the analytical form

\begin{equation}
p = b(\rho - \rho_0).
\end{equation}

The parameters $b$ and $\rho_0$ we take from \cite{Gondek-Rosinska2008} for EOS SQS B40 and SQS B60. SQS B60 is the softer one with a maximal mass below two solar masses, and SQS B40 is a stiffer one.

\section{Numerical results}

\subsection{Moment of inertia - compactness relations}

\subsubsection{Slowly rotating stars}

In this section we are investigating the universality in the moment of inertia--compactness  relations, suggested for the first time in \cite{Ravenhall1994}, and extensively studied  by Breu and Rezzolla  \cite{Breu2016} in GR. In this work we extend their study  for slowly rotating neutron and strange stars in STT and $\mathcal{R}^2$ gravity.

\begin{figure}[]
	\centering
	\includegraphics[width=0.45\textwidth]{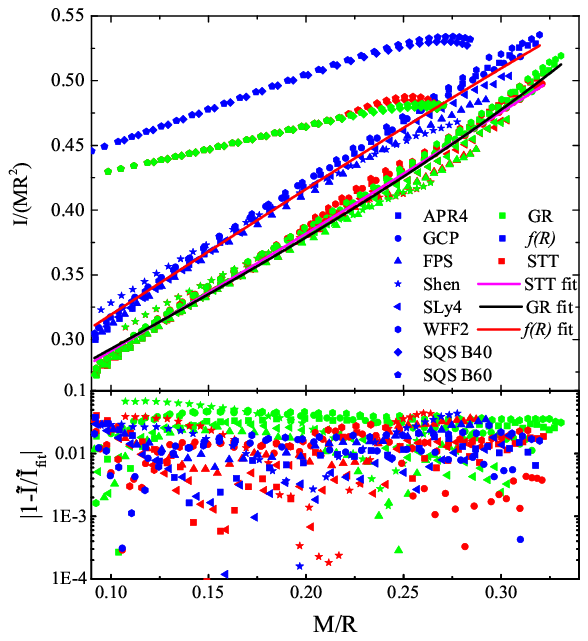}
	\includegraphics[width=0.45\textwidth]{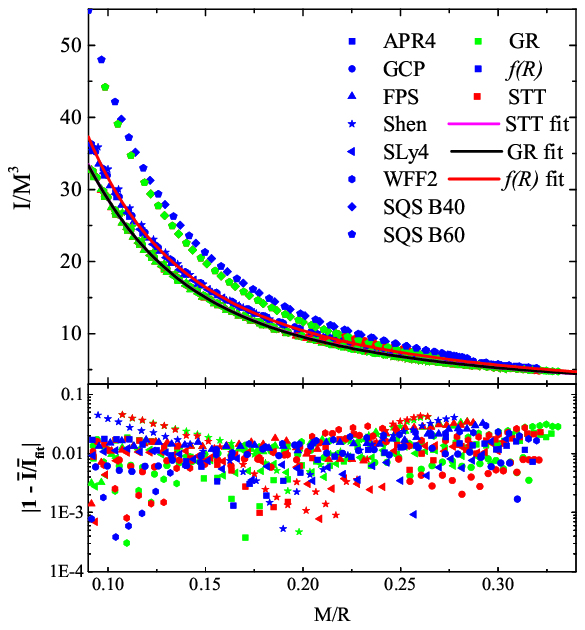}
	\caption{The results in slow rotation approximation are shown. The different EOS are marked with different symbols and the different theories -- with different colours. The plotted results are for neutron and strange stars in GR, in STT with $\beta = -4.5$ and in $\mathcal{R}^2$ gravity with $a = 10^4$. In the left panel  the results for $I/(MR^2)$ are plotted and in the right one -- for $I/M^3$. The results for neutron stars are fitted with polynomial fits and the deviations from universality $\left|1 - \tilde{I}/\tilde{I}_{\rm fit}\right|$ and $\left|1 - \bar{I}/\bar{I}_{\rm fit}\right|$ are plotted in the lower  panels of the corresponding graphs.}
	\label{Fig:slow}
\end{figure}

In the left panel of Fig. \ref{Fig:slow} we are plotting $ \tilde{I} \equiv I/(MR^3)$ as a function of the compactness. In green we plot the results for GR, in red are the STT results and in blue are the $f(\mathcal{R})$ ones.  The scalarized solutions in the STT theory are visible, but expectedly small. The $f(\mathcal{R})$ results are qualitatively the same as the GR ones, but they are shifted to higher values of $\tilde{I}$.

All the results for neutron stars show quite good universality within the theory, although for higher compactness the data points are more disperse. The strange stars case, on the other hand, is qualitatively different -- for small compactness the normalised moment of inertia  is  higher than the neutron star one, but for high values of $M/R$ the strange stars results for $\tilde{I}$ converge to the neutron star ones. We are fitting only the neutron star results, separately for the different theories, with a polynomial fit. The forth order polynomial, with excluded second and third order therms gives a small correction to the lineal fit which is a natural choice, considering the form of the results. The following form  is suggested for the first time by Lattimer and Schutz \cite{Lattimer2005} and recently used by Breu and Rezzolla \cite{Breu2016}:

\begin{equation} \label{eq:fit_tilde}
\tilde{I} = \tilde{a}_0 + \tilde{a}_1 \frac{M}{R} + \tilde{a}_2 \left(\frac{M}{R}\right)^4.
\end{equation}

The numerical values of the fitting coefficients are given in Table \ref{Tbl:IMRR}.
In the lower panel of the graph we are plotting the relative deviation of the data points from the fitting curve. The deviation we define as $|1 - \tilde{I}/\tilde{I}_{\rm fit}|$ and for the set of EOS we are using  it is below 10 \% for all theories studied here.

We are using eq. \ref{eq:fit_tilde} and the fitting coefficients to make a rough numerical estimate for the difference between GR and $\mathcal{R}^2$ gravity. For compactness $0.2$ the difference is about 10\%. Due to the small difference between the GR and the STT results, we do not do this evaluation for the STT ones, but if one is interested  in the numbers, the computations can be performed easily by using eq. \ref{eq:fit_tilde} and Table \ref{Tbl:IMRR}.

The soft  FPS EOS  does not show any significant deviations from universality. EOS  Shen, however, separates from the other models for low compactness leading to higher values for the normalised moment of inertia. This is valid for the GR results as well as for the $f(\mathcal{R})$ ones, although the difference is small.

In the right panel of Fig. \ref{Fig:slow} we are using a different normalisation for the moment of inertia, namely $\bar{I} \equiv I/M^3$. In this case the normalised moment of inertia decreases with the increase of the compactness. The scalarized branch in the STT case is visible. The difference between the GR and the $\mathcal{R}^2$ gravity results is significantly decreased and  both theories  coincide for models with maximal compactness. For this normalisation of the moment of inertia the strange star results separate from the ones for neutron stars  too. The strange star  results look qualitatively the same with the neutron stars one, but again for small compactness the normalised moment of inertia has larger values and converges to the neutron star one for maximal $M/R$. We tested different forms of the fitting function for this case. Fitting functions in powers of $M/R$ are non-converging, therefore,  inconvenient in this case. The extensive study of different polynomial functions of the reverse compactness showed that the forth order therm slightly decreases the deviation from EOS universality with respect to the third order polynomial fit. The absence of the linear therm would lead to bigger deviations for maximal compactness. Adding a free therm to the present fitting function decrease slightly more the deviation for models with maximal compactness. For consistency we are using the polynomial fit, suggested and motivated in \cite{Breu2016}, in the following form:

\begin{equation} \label{eq:fit_bar}
\bar{I} = \bar{a}_1 \left(\frac{M}{R}\right)^{-1} + \bar{a}_2 \left(\frac{M}{R}\right)^{-2} + \bar{a}_3 \left(\frac{M}{R}\right)^{-3} + \bar{a}_4 \left(\frac{M}{R}\right)^{-4}.
\end{equation}
The numerical values for the fitting coefficients are given in Table \ref{Tbl:IMMM}. Again we are using the fitting function and the coefficients from the table to estimate the rough deviation of the $\mathcal{R}^2$ gravity results from the GR ones. The deviation is about 11 \% for compactness $0.1$ and decreases to about 7 \% for $M/R = 0.3$.

In the lower panel of the graph we are plotting the deviation from EOS universality, $|1 - \bar{I}/\bar{I}_{\rm fit}|$, for neutron stars. Although it looks like this normalisation leads to better universality, this impression is due to the much bigger domain for the normalised moment of inertia as  the results show.
The deviation is about 5 \% for our set of EOS, and it is comparable with the previous case.
EOS FPS and EOS Shen do not break the universality and the results does not show  significantly higher deviations from the fitting curve for these two EOS compared to the rest of the EOS set.

\subsubsection{Rapidly rotating stars}

In this section we expand our study to the case of rapidly rotating neutron and strange stars. In the literature different constant rotational parameters along the sequences of models have been studied \cite{Doneva2014, Pappas2014, Chakrabarti2014, Doneva2015, Doneva2014a}. In \cite{Doneva2014} it has been shown that if we are using constant angular velocity $\Omega$, the $\bar{I} - \bar{Q}$ universality is lost. In order to preserve the universality we are using the dimensionless  angular momentum $j = J/M^2$ as a fixed parameter, similar to \cite{Chakrabarti2014}. The values for $j$ we use are $j = 0.2, j = 0.4,$ and $j = 0.6$. For small masses the $j = 0.2$ case overlap with the angular velocity range in which the slow rotation approximation should be accurate. With the increase of the mass, however, the angular velocity increases out of the validity range for the slow rotation approximation. Therefore the models obtained in the slow rotation approximation will not be the same as the ones for $j = 0.2$ which motivate us to use such small value of $j$.

\begin{figure}[]
	\centering
	\includegraphics[width=0.45\textwidth]{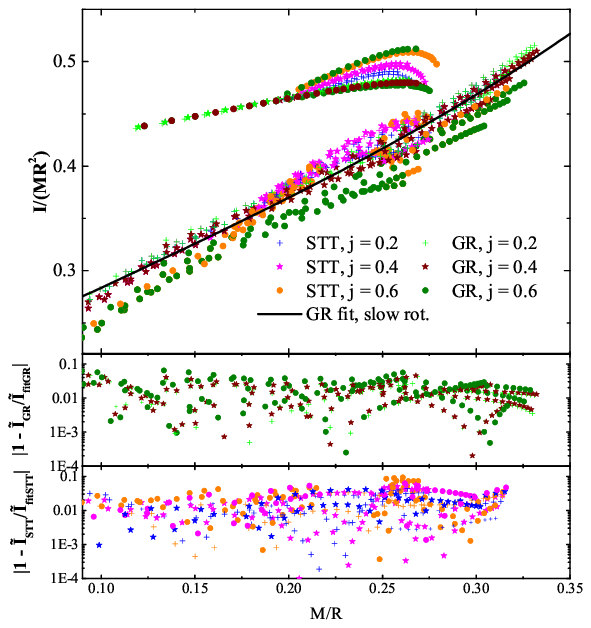}
	\includegraphics[width=0.45\textwidth]{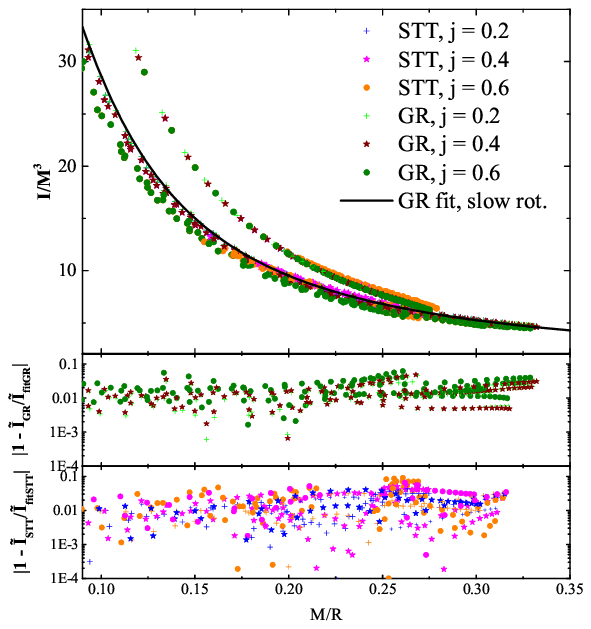}
	\caption{The results for rapidly rotating stars are shown. The different theories and the different rotational rates are marked with different symbols and colours. The plotted results are for neutron and strange stars in GR and in STT with $\beta = -4.5$. In the left panel  the results for $I/(MR^2)$ are plotted and in the right one -- for $I/M^3$. The results for neutron stars are fitted with polynomial fits and the deviations from universality $\left|1 - \tilde{I}/\tilde{I}_{\rm fit}\right|$ and $\left|1 - \bar{I}/\bar{I}_{\rm fit}\right|$ are plotted in the lower  panels of the corresponding graphs. }
	\label{Fig:rapid_STT}
\end{figure}

In Fig. \ref{Fig:rapid_STT} we are plotting the GR and the STT results. The different theories and the different rotational rates are marked with different symbols and colours. The fitting curve for the GR case in slow rotation approximation is plotted in black continuous line. Because of the minor difference between the GR and the STT fits we are skipping the latter one.
In the left panel the results for $\tilde{I}$ are plotted. They are qualitatively the same as in the slowly rotating case. The deviations of the scalarized solutions from the GR ones  increase with the increase of the angular momentum of the star. The increase of $j$ shifts the results to lower values of $\tilde{I}$.
It is visible that the case with $j = 0.2$ coincides with the slow rotation approximation and the difference  even with the $j = 0.6$ case is not very large.

The results are fitted with the polynomial, given by eq. (\ref{eq:fit_tilde}) and the fitting coefficients are given in Table \ref{Tbl:IMRR}.
The fitting curves are not plotted so as to keep the graph maximally clear. In the lower panel the relative deviations between the data points and the fitting curves are plotted.  In GR and in STT  the maximal deviation from EOS universality is below 10 \% for all values of $j$.

In the right panel the same results in terms of $\bar{I}$ are plotted. The case of $j = 0.2$ matches very well with the slow rotation fitting curve. The increase of $j$ leads to decrease of $\tilde{I}$ but in general the difference from the slow rotation is small. The results are fitted with eq. (\ref{eq:fit_bar}) and the fitting coefficients are given in Table \ref{Tbl:IMMM}. In the lower sections  the deviations from EOS universality are plotted. The values are below 10 \%.

\begin{figure}[]
	\centering
	\includegraphics[width=0.45\textwidth]{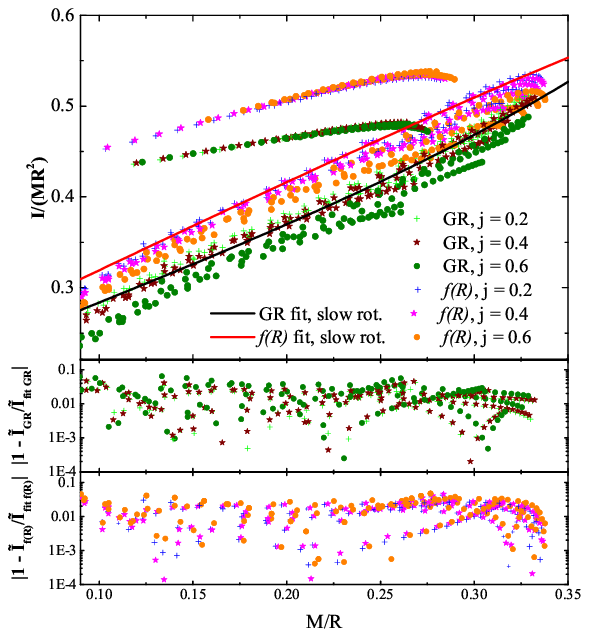}
	\includegraphics[width=0.45\textwidth]{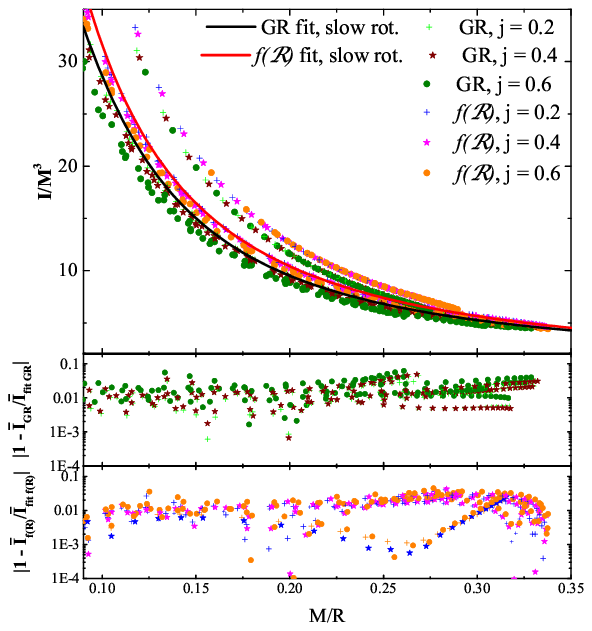}
	\caption{The results for rapidly rotating stars are shown. The different theories and the different rotational rates are marked with different symbols and colours. The plotted results are for neutron and strange stars in GR and in $\mathcal{R}^2$ gravity with $ a = 10^4$. In the left panel the results for $I/(MR^2)$ are plotted and in the right one -- for $I/M^3$. The results for neutron stars are fitted with polynomial fits and the deviations from universality $\left|1 - \tilde{I}/\tilde{I}_{\rm fit}\right|$ and $\left|1 - \bar{I}/\bar{I}_{\rm fit}\right|$ are plotted in the lower  panels in the corresponding graphs. }
	\label{Fig:rapid_fR}
\end{figure}

In Fig. \ref{Fig:rapid_fR} we are plotting the results for GR and for $\mathcal{R}^2$ gravity. Just like in the figures above, in the left panel the results for $\tilde{I}$ are plotted  and in the right one -- for $\bar{I}$. The results for $\tilde{I}$ in both theories  coincide with the corresponding slow rotation fitting curves for $j = 0.2$.  As it was commented above, for GR the increase of the normalised angular momentum leads to a decrease of the normalised moment of inertia. This is the case for the $\mathcal{R}^2$ gravity results too. Because of that the results for the maximal adopted value of $j$ ($j = 0.6$) in the $f(\mathcal{R})$ theory gets close to the GR results for $j = 0.2$ and for maximal compactness they partially overlap.

At this point an interesting observation on the strange star results should be made. In the figure one can see that the change of $\tilde {I}$ due to the rotation is marginal for  both of the examined quark EOS.

In the right panel of Fig. \ref{Fig:rapid_fR} the results for $\bar{I}$ and the corresponding slow rotation fitting curves are plotted. Expectedly the slow rotation fits coincide with the $j = 0.2$ case in  both theories.  Due to the smaller deviation of the results between the theories compared to the left panel, the effect of converging of the results with high value of $j$ in $\mathcal{R}^2$ gravity  to the GR results with small value of $j$ is even more pronounced.

We are fitting the GR and the $f(\mathcal{R})$ results with the polynomial given by eq. (\ref{eq:fit_tilde}) and eq. (\ref{eq:fit_bar}) in the left and  the right panel respectively. The fitting coefficients are given in Tables \ref{Tbl:IMRR} and \ref{Tbl:IMMM}.

In both panels of the figure we are plotting the deviations from universality for GR and for $R^2$ gravity. In all cases the deviations are below 10 \%. At this point a comment concerning the results for rapid rotation in \cite{Breu2016} should be made. In that paper the authors report deviations of about 20 \% for the $j = 0.6$ case. The difference with our paper most probably, comes from  the different set and the smaller number of EOS we are using in our study. However,  our main goal is to study the examined relations in alternative theories of gravity and to compare them with the GR case. The results show similar behaviour of the relations in all examined theories.

\begin{table}[h]
\begin{tabular}{cccccccc}
\hline
\quad & $\tilde{a}_0$ & $\tilde{a}_1$ & $\tilde{a}_2$ & $\chi^2_{\rm red}$\\
\hline
\multicolumn{2}{c}{GR} \\
\cline{1-2}
slow. rot.     & 0.210    & 0.824     & 2.480    &  $4.650 \times 10^{-5}$  \\
j = 0.2          & 0.211    & 0.788     & 3.135     &  $4.314 \times 10^{-5}$  \\
j = 0.4          & 0.200   & 0.823    & 2.469    &  $4.595 \times 10^{-5}$  \\
j = 0.6          & 0.176    & 0.839     & 2.393     &  $7.682 \times 10^{-5}$  \\
\multicolumn{2}{c}{STT} \\
\cline{1-2}
slow. rot.     & 0.201    & 0.897    &   0.603   & $4.750 \times 10^{-5}$   \\
j = 0.2          & 0.196    & 0.909    &   0.224   & $5.633 \times 10^{-5}$   \\
j = 0.4          & 0.171   & 1.055    &  $-2.985 $   & $1.987 \times 10^{-4}$   \\
j = 0.6          & 0.127   & 1.256     &  $-7.190$   & $1.794 \times 10^{-4}$   \\
\multicolumn{2}{c}{$f(\mathcal{R})$} \\
\cline{1-2}
slow. rot.     & 0.221   & 0.981     &  $ -0.755$    &  $4.914 \times10^{-5} $  \\
j = 0.2          & 0.216   & 0.989     &   $-1.255$    &  $5.562 \times 10^{-5}$  \\
j = 0.4          & 0.208   & 1.011     &   $-1.748$    &  $6.026 \times 10^{-5}$  \\
j = 0.6          & 0.201   & 0.998     &   $-2.143$    &  $8.097 \times 10^{-5}$  \\
\hline
\end{tabular}
\caption{The fitting coefficients for the fit given by eq. (\ref{eq:fit_tilde}). The results are given for the slowly rotating case and   rapidly rotating case with $j = 0.2$, $j = 0.4$, $j = 0.6$, for the GR case, followed by the STT case and by the $f(\mathcal{R})$ case. In the last column  the reduced $\chi ^2$ values are printed.  }
\label{Tbl:IMRR}
\end{table}

\begin{table}[h]
\begin{tabular}{cccccccc}
\hline
\quad & $\bar{a}_1$ & $\bar{a}_2$ & $\bar{a}_3$ & $\bar{a}_4$ & $\chi^2_{\rm red}$\\
\hline
\multicolumn{2}{c}{GR} \\
\cline{1-2}
slow. rot.     & 1.165    & 0.0538    &   0.0259 &  $- 0.00144$ &  0.0413  \\
j = 0.2          & 1.077    & 0.100     &   0.0172 & $-9.830 \times 10^{-4} $ &  0.0219  \\
j = 0.4          & 1.024    & 0.123      &    0.0129 & $-7.985 \times 10^{-4}$   &  0.0249  \\
j = 0.6          & 0.943    & 0.143      &   0.00714 & $-5.539 \times 10^{-4}$    &  0.0507  \\
\multicolumn{2}{c}{STT} \\
\cline{1-2}
slow. rot.     & 1.057    & 0.110    &  0.0173    & $-0.00103$ & 0.0420  \\
j = 0.2          & 0.884   & 0.197    &   0.00270    & $-3.254 \times 10^{-4}$ & 0.0183  \\
j = 0.4          & 0.664     & 0.313     &   $-0.0163$    & $ 5.647 \times 10^{-4}$ & 0.0272  \\
j = 0.6          & 0.257     & 0.513     &   $-0.0492$   & 0.00202 & 0.0655  \\
\multicolumn{2}{c}{$f(\mathcal{R})$} \\
\cline{1-2}
slow. rot.     & 0.941    & 0.214      &   0.00521     & $-4.412 \times 10^{-4}$ & 0.0633  \\
j = 0.2          & 0.853    & 0.243    &  0.00190    & $-3.841 \times 10^{-4}$ & 0.0190  \\
j = 0.4          & 0.805   & 0.264      &  $-0.00201$    & $-1.978 \times 10^{-4}$ & 0.0168  \\
j = 0.6          & 0.725    & 0.280      &  $-0.00407$     & $-1.706 \times 10^{-4}$ & 0.0286  \\
\hline
\end{tabular}
\caption{The fitting coefficients for the fit given by eq. (\ref{eq:fit_bar}). The results are given for the slowly rotating case and   rapidly rotating case with $j = 0.2$, $j = 0.4$, $j = 0.6$, for the GR case, followed by the STT case and by the $f(\mathcal{R})$ case. In the last column  the reduced $\chi ^2$ values are printed.    }
\label{Tbl:IMMM}
\end{table}

\subsection{Universality for maximal masses}

In this section we are investigating the maximal  masses for sequences with fixed angular momentum, $(\partial{M}/\partial{\rho_c})_J = 0$, or as they are called in \cite{Breu2016} -- models with "critical masses". As shown in \cite{Takami2011} the models with critical masses are unstable (see also \cite{Breu2016}). The situation in the alternative theories is much more complicated but it is natural to expect that the models under considerations are also unstable  in scalar-tensor and $f(\mathcal{R})$ theories. The results for all  values of the angular momentum are calculated with the above mentioned modification of the $RNS$ code.

\begin{figure}[]
	\centering
	\includegraphics[width=0.45\textwidth]{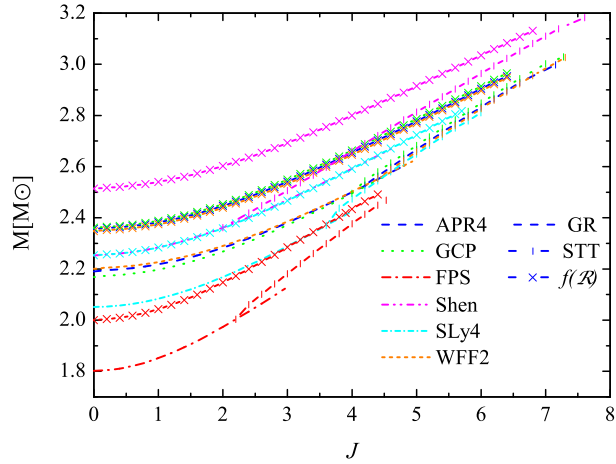}
	\caption{The maximal masses of sequences with fixed angular momentum as a function of the angular momentum. The different EOS are marked with different colours and patterns. The GR results are noncontinuous lines without any symbols, the STT ones and the $f(\mathcal{R})$ ones are marked with different symbols.}
	\label{Fig:M_J}
\end{figure}

We are interested in
 the relations between the properly normalised mass and  angular momentum of those unstable models.
In Fig. \ref{Fig:M_J}  the maximal mass as a function of the dimensionless angular momentum $J$ is plotted. The connection between the dimensional and the dimensionless angular momentum is the following $J \rightarrow J/(GM_{\odot}/c)$. 
The results for different EOS are marked with different colours and patterns. The GR results are in noncontinuous lines without any symbols, the STT ones and the $f(\mathcal{R})$ ones are marked with different symbols.
The results in $f(\mathcal{R})$ theories  are generally with higher masses than the corresponding GR and STT results, but due to the high dispersion, caused by the EOS uncertainty,  the results for all theories overlap.

\begin{figure}[]
	\centering
	\includegraphics[width=0.45\textwidth]{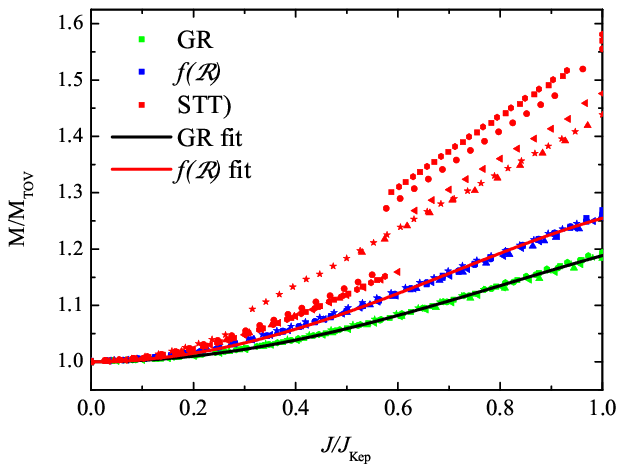}
	\includegraphics[width=0.45\textwidth]{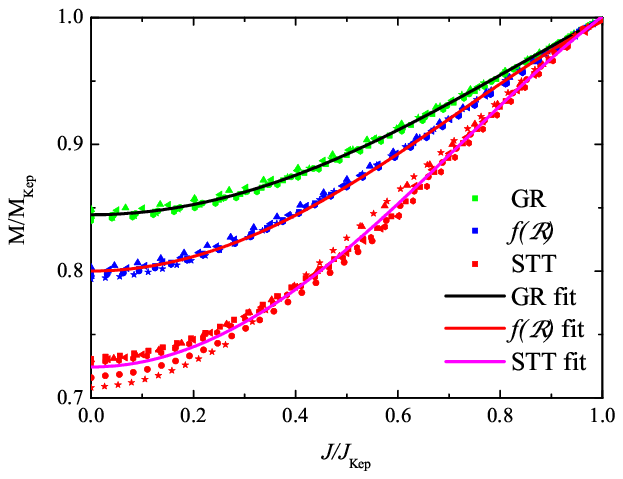}
	\caption{The maximal mass normalised to the corresponding TOV mass in the left panel and normalised to the maximal Keplerian  mass as a function of the angular momentum, normalised to the maximal Keplerian one. The results for neutron stars in the three examined theories are plotted in  both panels. }
	\label{Fig:Mmax}
\end{figure}

In Fig. \ref{Fig:M_J} one can see that the maximal angular momentum is considerably higher than the one in the GR case in the examined alternative theories. For strange stars $J$ is generally higher than for neutron stars  and  the difference becomes significant in the alternative theories. If we plot the strange star results in  Fig. \ref{Fig:M_J}, the rest of the plots will get very unclear. When we combine that with the fact that the strange stars solutions deviate significantly from the examined universality, we chose not to plot the strange stars results in Figs. \ref{Fig:M_J} and \ref{Fig:Mmax}.

An additional comment on how we are determining the maximal mass in the STT case, for a specific EOS, should be made. The GR case is a solution to the STT for the scalar field $\varphi = 0$. For a non-vanishing scalar field a scalarization of the solution occurs only for a specific interval of central energy densities, and for values not in this interval, the GR solutions are obtained.  The difference between the scalarized and the GR solutions, generally small for static and slowly rotating stars,  increases considerably for higher angular momentum \cite{Doneva2013}. Because of that the maximal masses sequence we are interested in will be combination of GR maximal masses for slowly rotating stars and maximal masses from the scalarized branch for rapidly rotating stars. At some value of $J$ the maximal mass in the scalarized branch get bigger than the maximal mass in the GR sequence.  From that point on we start to plot the maximal mass for the scalarized solution. In the figure the point at which the GR and the STT solutions separate is clearly distinguishable.

In Fig. \ref{Fig:Mmax} we are studying  different normalisations so that the equation of state independent relations overcome the dispersion in Fig. \ref{Fig:M_J}. In the left panel the maximal mass is normalised to the mass of the  non-rotating models with the same central energy density marked with  $M_{TOV}$. The angular momentum $J$ is normalised to the maximal Keplerian angular momentum for the EOS, $J/J_{\rm Kep}$. The colour and symbol notation is as in the graphs above. Considering the normalizations, the results for all EOS expectedly coincide for static and slowly rotating models. The results in $\mathcal{R}^2$ gravity are with higher values for $M/M_{TOV}$ compared to the GR ones.  For both theories there is good universality. The results are fitted with a polynomial fit of the form \cite{Breu2016}:

\begin{equation} \label{eq:Mmax_tov}
\frac{M}{M_{\rm TOV}} = 1 + a_1\left(\frac{J}{J_{\rm Kep}}\right)^2 + a_2\left(\frac{J}{J_{\rm Kep}}\right)^4.
\end{equation}
The coefficients of the fit are given in Table \ref{Tbl:max}.

However, the STT results in Fig. \ref{Fig:Mmax} are  much more interesting. The angular momentum, for the entire sequence of models, is normalised to the maximal one, determined for the scalarized solution. In Fig. \ref{Fig:M_J} one can see, that the increase of $J$ is substantial compared to the GR one. This is why the deviations from GR occur for small values of $J/J_{\rm Kep}$.  At the point where the  maximal mass moves to the scalarized solution there is a discontinuity in the $M/M_{\rm TOV}$ sequence in te left panel of Fig. \ref{Fig:Mmax}. The explanation is straightforward: at the value of $J/J_{\rm Kep}$ at which the mass maximum is moved to the scalarized branch,  there is a change, from high central energy densities to low ones due to the fact that the scalarized solution occurs for lower central energy densities than the maximal mass in GR. Therefore there is a significant decrease of the mass of the corresponding static models. This change has a serious effect on the normalisation as can be seen on the graph. It is interesting to  mention that the universality in the STT case is lost and the dispersion of the results is comparable to the one caused by the EOS uncertainty. Due to the discontinuity we do not fit the STT set of data.

We explored a different normalisation that seems more natural to us.
 In the right panel of Fig. \ref{Fig:Mmax}  we are normalising the maximal mass to the maximal Keplerian one for the EOS, $M/M_{\rm Kep}$.

The deviation between the theories is maximal for static models and decreases with the increase of the normalised angular momentum, which occurs naturally  because of the normalisation. The $M/M_{\rm Kep}$ values decrease in  $\mathcal{R}^2$ gravity, compared to the GR ones, and the lowest ones are in the STT case. Just like in the left panel, the results for GR and $f(\mathcal{R})$ gravity show good universality.

The STT case deserves a special attention. Due to the change of the normalisation the dispersion of the results is notably decreased for the whole set of STT data. The discussed above discontinuity is absent. The normalisation restores the universality up to a good extent, so we are fitting this case too. The polynomial fit we are using is the flowing:

\begin{equation} \label{eq:Mmax_kep}
\frac{M}{M_{\rm Kep}} = a_0 + a_1\left(\frac{J}{J_{\rm Kep}}\right)^2 + a_2\left(\frac{J}{J_{\rm Kep}}\right)^4,
\end{equation}
and the coefficients are given in Table \ref{Tbl:max}.

\begin{table}[h]
\begin{tabular}{cccccccc}
\hline
\quad & $a_0$ & $a_1$ & $a_2$ & $\chi^2_{\rm red}$\\
\hline
\multicolumn{1}{c}{$\frac{M}{M_{TOV}}\left(\frac{J}{J_{Kep}}\right)$ } \\
\cline{1-2}
   GR & 1  & 0.251      &  $ -0.0626$    &  0.729$\times10^{-5}$  \\
$f(\mathcal{R})$     & 1  & 0.382     &  $ -0.127$    & 1.394 $\times 10^{-5}$  \\
STT   &  -- & -- & -- & --\\

\multicolumn{1}{c}{$\frac{M}{M_{Kep}}\left(\frac{J}{J_{Kep}}\right)$} \\
\cline{1-2}
GR  & 0.844    & 0.203      &   $-0.0484$    &  0.422 $\times 10^{-5}$  \\
$f(\mathcal{R})$  & 0.800    & 0.290    &   $-0.0933$   & 0.533 $\times 10^{-5}$  \\
STT & 0.724  & 0.403  & $-0.126$  & 3.515 $\times 10^{-5}$\\
\hline
\end{tabular}
\caption{The fitting coefficients for the maximal mass for the fits (\ref{eq:Mmax_tov}) and (\ref{eq:Mmax_kep}). In the last column the reduced $\chi ^2$ values are given.}
\label{Tbl:max}
\end{table}

\section{Conclusions}

Universal relations between the moment of inertia, the mass and the radius of a neutron star  are very appealing. The mass and the moment of inertia can be measured with good accuracy, while optical measurement of the radius depends on different factors, like the redshift, the distance to the star, the effect of the atmosphere, the absorption in the interstellar space, etc. \cite{Lattimer2005}.  Because of the large uncertainty in the EOS such relations allow us to determine uniquely the star's parameters.     They provide us with an accurate tool for testing GR and the alternative theories.

$M, I$ and $R$ are measurable at this point, which makes EOS independent relations between these parameters more appealing than some of the most popular in the literature.
The $I$-Love-$Q$ relations (or ones between higher multipole moments) and relations concerning oscillation frequencies demand higher multipole moment or gravitational waves frequencies to be measured. On the other hand measuring $M$ and $I$ with a good accuracy for  binary pulsar systems can be easily achieved.

Our study shows that using  $I/(MR^2)$ or $I/M^3$ as a function of the compactness leads to quite good universality in agreement with \cite{Breu2016}. In the case of slowly rotating as well as rapidly rotating neutron stars the deviations, for our set of EOS, are not higher than 10 \% for both studied normalisations. The results for strange stars, on the other hand,  give a hint for  universality, but more quark EOS should be tested in the future. If it turns out that there is a large deviation between strange stars universal relations and neutrons stars ones, it would allow testing the nature of the compact objects.

The universal relations can be used not only to determine the star's parameters but also to  restrict the theory itself. In general the deviations in the results due to the uncertainty in the EOS are comparable or even higher than the deviations due to the different gravitational theory. By using EOS independent relations we can concentrate on the latter one, without being affected significantly by the former one. In this study we are using two viable alternative theories of gravity. In all examined cases the deviation from universality is below 10 \%, for our set of EOS, like in the GR case. The results for STT do not show significant deviations from the GR ones, not even for the rapidly rotating models.  These relations are in some sense even theory independent. This is not the case for rapidly rotating strange stars when using $I/(MR^2)$, where major deviations from GR in the STT regime can be observed.

The $\mathcal{R}^2$ gravity is the more interesting case. We know that the maximal deviations for the moment of inertia can be of the order of 40 \% for the EOS studied in  \cite{Staykov2014}. In the examined relations the $I/(MR ^2)$ results show deviation from GR of about 10 \% for the whole interval of compactnesses. In the $I/M^3$ case the maximal deviation is above 10 \% and decrease  with the increase of the compactness. The rapidly rotating case is more interesting to be discussed. With the increase of $j$ the $\mathcal{R}^2$ results get closer to the slowly rotating GR results. This leads us to the conclusion that the slowly rotating case (or fit) can be used with good approximation for the description not only of rapidly rotating models in GR but also for rapidly rotating models in $\mathcal{R}^2$  gravity. We should stress that in this study we have used the maximal value for the free parameter of the theory allowed by the observations . Using lower values of the parameter will give even smaller deviation from GR. Hence, these relations are not only EOS independent but for the most of the domain of the chosen $f(\mathcal{R})$ theory they  will be theory independent too.

It was interesting to show that universal relations can be found for maximal mass models like the ones in Fig. \ref{Fig:Mmax}.  These results are a good representation of how sensitive the relations can be to the normalisations in different theories.

\section*{Acknowledgements}

The authors would like to thank K. Kokkotas and L. Rezzolla for the discussions and the useful suggestions. DD would like to thank the European Social Fund and the Ministry Of Science, Research and the Arts Baden-Württemberg for the support. KS and  SY would like to thank the Research Group Linkage Programme of the Alexander von Humboldt Foundation for the support.
The support by the Bulgarian NSF Grant DFNI T02/6 is gratefully acknowledged.


\bibliography{references}

\end{document}